\documentstyle[pre,aps,multicol,epsf]{revtex}
\def\lsim{\mathrel{\hbox{\rlap{\hbox{\lower4pt\hbox{$\sim$}}}\hbox{$<$}}}}
\def\gsim{\mathrel{\hbox{\rlap{\hbox{\lower4pt\hbox{$\sim$}}}\hbox{$>$}}}}
\newcommand\med[1]{\langle {#1}\rangle}
\begin{document}
\draft
\title{\bf{Sequence randomness and polymer collapse transitions.}}
\author{Pietro Monari $^1$, Attilio L. Stella ${^{1,2}}$,
Carlo Vanderzande $^3$ and Enzo Orlandini$^1$\\}
\address{$^1$ INFM-Dipartimento di Fisica, Universit\`a
di Padova, I-35131 Padova, Italy\\
$^2$ The Abdus Salam ICTP, P.O. Box 586, 34100 Trieste, Italy\\
Sezione INFN, Universit\`a di Padova, 35100 Padova, Italy\\
$^3$ Departement WNI, Limburgs Universitair Centrum, 3590 Diepenbeek, Belgium}
\maketitle

\begin{abstract}

Contrary to expectations based on Harris' criterion,
chain disorder with frustration can modify the universality class of scaling 
at the theta transition of heteropolymers. This is shown for a
model with random two-body potentials in $2D$ on the basis of exact
enumeration and accurate Monte Carlo results. When frustration grows beyond a 
certain finite threshold, the temperature below which disorder becomes
relevant coincides with the theta one and scaling exponents definitely 
start deviating from those valid for homopolymers.

\end{abstract}
\pacs{05.70.Jk,64.60.-i,64.60.Kw,36.20.-r}
%-----------------------------------------------------------
% Document
%-----------------------------------------------------------

\begin{multicols}{2}
\narrowtext
In recent years, work on polyampholytes and on biomolecules, like
proteins, has focused much
attention on the conformational properties of inhomogeneous 
polymeric chains (heteropolymers)\cite{garel}.
Some of these properties, like protein folding, are expected
to be determined by the specific sequence of different monomers constituting 
the chain\cite{chan}. In this context, particularly relevant is
a statistical point of view, according to which the behavior of large
ensembles of different sequences is globally tested with philosophy and
methods  of the physics of quenched disordered systems\cite{bryngelson}. 
Such studies are intended to provide
information on the conditions under which specificity becomes
important (i.e., disorder becomes relevant) and to give 
global descriptions of its    
possible effects.

The collapse from swollen
to compact globular state of a long macromolecular chain
is by now well understood
in the case of homopolymers. At temperature $T$, a chain with $N>>1$ 
equal monomers has an
average radius of gyration $R_N\propto N^{\nu(T)}$. $\nu(T)$
is equal to the exponent of a self--repelling chain (SAW)\cite{vanderzande}
for all temperatures $T>T_{\theta}$, and to
$1/d$ for $T<T_{\theta}$, as appropriate
for a compact object. $\nu(T_{\theta})=\nu_{\theta}$ has a distinct,
intermediate value, known both
in $3D$ ($1/2$)\cite{degennes} and in $2D$ ($4/7$)\cite{duplantier}.
This transition is triggered by attractive
interactions between monomers as $T$ decreases. 
In the case of heteropolymers, 
%like proteins
%is temperature is lowered, 
the theta transition involves the same
sequence of regimes and
preludes to the folding phenomenon, for which specificity
is surely of key importance. Thus, one can wonder if chain disorder could 
substantially affect polymer behavior already at the onset of 
theta collapse, or even at higher $T$.
 
Unlike random environment disorder for a homopolymer,
if amounting to a small perturbation,
inhomogeneities in the structure of the chain should in general
be expected to be irrelevant and not able to affect the ordered system's
behavior, as far as universal scaling is concerned.
This is suggested by
Harris' criterion\cite{harris} as we discuss below. In fact, 
contrary to previous conjectures\cite{golding}, recent work on a model of
randomly charged polymers in $2D$ and $3D$,
has shown that disorder does not change the universality class of the
theta transition\cite{monari}, consistent with the most simple 
scenario one could infer based on application of the above criterion. 
However, in this Letter we give evidence
that sufficiently large amounts of chain disorder and frustration can modify
the heteropolymer theta behavior,
with respect to the homopolymer one. Thus, in such situations, chain 
specificity 
becomes a key ingredient in determining the universality class of the 
theta point. Quite remarkably, this point seems to fall right at the upper 
limit of temperatures for which disorder plays a relevant role.

As heteropolymer model we consider here an $N$-step SAW on square
lattice. On each lattice site visited by the walk sits a monomer.
Monomers $i$ and $j$ ($0\leq i,j\leq N$) 
which are not consecutive along the chain
($i\neq j$, $|j-i|\neq 1$) and occupy nearest neighbor lattice sites
in a given configuration, feel an attractive potential $V_{ij}$
( $i$ and $j$ constitute
a contact indicated by $\med{ij}$).
$V_{ij}$ is random with probability distribution
$P(V_{ij})=p\delta(V_{ij}+V)+(1-p)\delta(V_{ij}-V)$, ($V<0$).
Its values are assigned
independently to each pair of monomers along the sequence.
Models of this kind were already used for
proteins\cite{bryngelson}. We choose it here because,
since disorder is associated to
monomer pairs, rather than to individual monomers, the annealed partition 
can be easily mapped into a well defined effective homopolymer problem. 

One has to compute
free energy and other thermal averages for each possible potential
arrangement along the sequence. Results have then to be further
averaged over disorder. In addition one wants to establish how important
fluctuations due to different disorder configurations are in the evaluation
of the final results. A typical case is that of the free energy: the quenched
quantity is the disorder average of the logarithm of the partition function, 
$Z_N\{V\}=
\sum_{\omega}exp(-\sum_{\med{ij} \in\omega}V_{ij}/T)$,
where the number of steps
of all the chain configurations $\omega$ is implicitly assumed equal to $N$.
If the distribution of $Z_N$ values is sharply peaked around its 
disorder average, $\overline{Z_N}=\Sigma_{\{V\}}\Pi_{\med{ij}}
P(V_{ij})Z_N\{V\}$, we have 
$ln(\overline{Z_N})=\overline{ln(Z_N)}$ for $N\to \infty$, and
disorder plays no role, i.e. annealed and quenched free energies,
respectively, are identical.

Since the potentials for different $\med{ij}$ are independent random variables,
the annealed problem reduces to a standard homopolymer one with an effective
attractive interaction $-T\log(\overline{exp(-V_{ij}/T)}$.
Imagine now to
perturb with a slight disorder ($p\sim 0$) the attractive homopolymer situation
($p=1$, $V_{ij}=-V$). The possible relevance of disorder can then be
discussed by looking at $n$ replicas of the SAW, for which the average 
partition can be put in the form:
\begin{eqnarray}
\lefteqn{\overline{Z_N^n}=
\sum_{\omega_{\alpha},\alpha=1,..n} \exp
\left[ -\frac{1}{T}\sum_{\beta}\sum_{\med{ij}\in 
\omega_{\beta}}\overline{V_{ij}}+ \right.}
\nonumber \\
& & \left.
- \frac{1}{2T^2}\sum_{\gamma,\delta}\sum_{\med{kl}\in \omega_{\gamma}}
\sum_{\med{mn} \in\omega_{\delta}}
\overline{\delta V_{kl} \delta V_{mn}}+..)   
\right ]
\end{eqnarray}
where a cumulant expansion has been used for disorder averages, and
$\overline{V_{ij}}=V(1-2p)$, while $\overline{\delta V_{ij}\delta V_{kl}}=
4p(1-p) \delta_{i,l} \delta_{j,k}$. Thus, if $\overline{V_{ij}}/T$ is
fixed to the value appropriate for the renormalization group fixed point
of a homopolymer at the theta transition, according to eq.1
the leading perturbation to
this fixed point is given by an operator proportional to the number
$I(\omega_{\gamma},\omega_{\delta}$)
of distinct contacts common to two replicas
in configurations $\omega_{\gamma}$ and $\omega_{\delta}$\cite{note}. 
The relevance or irrelevance of the disorder perturbation depends on 
whether, for the two replicas, the average of $I$ grows with $N$, or not.
By exact enumeration we studied this average for two 
replicas of up to $19$ steps without mutual interactions, 
and verified to high precision that indeed, at the theta
point, or even at lower temperatures, it saturates to a
constant for growing $N$. 
This implies irrelevance and could lead to expect that
also finite amounts of disorder would not be sufficient to subtract the
theta transition from the control of the homopolymer fixed point.
In such a scenario the border line temperature, $T_d$, below which
disorder becomes possibly relevant, should always satisfy $T_d<T_{\theta}$
strictly.

To determine $T_d$ is highly nontrivial. A straightforward strategy
could consist in extrapolating the ratio $\overline{Z_N^2}/{\overline{
Z_N}}^2$ to $N\to \infty$. According to general
theorems\cite{cook}, if this ratio tends to some finite $B\geq 1$, the 
annealed
free energy should be obtained for a fraction of all sequences summing up
to a probability $\geq 1/B$. Thus, if $B=1$, quenched
and annealed problems must coincide. $B$ is in principle adequate
only to establish an upper bound on $T_d$. In addition, since quantities like
$\overline{Z_N}$ are wildly diverging and sensibly oscillating 
for increasing $N$, $B$ estimates are problematic and we have
to use a different strategy\cite{monari}.
Besides $\nu$, also entropic exponents characterize SAW scaling.
For example, one expects $\overline{Z_N} \sim N^{\gamma_a-1} K_a^{-N}$
for $N \to \infty$\cite{vanderzande}. 
$\gamma_a$ must take on the distinct
values (appropriate for homopolymers)
$43/32$ and $8/7$, for $T>T_{a\theta}$ and
$T=T_{a\theta}$\cite{vanderzande,duplantier}, 
respectively, where $T_{a\theta}$ indicates the theta temperature of the
annealed problem. For $T<T_{a\theta}$ the precise value of $\gamma_a$
is still debated and may be non-universal, depending on lattice and
boundary conditions\cite{compact}.
If the polymer
has one end fixed on an impenetrable boundary in semi--infinite
geometry, the behavior of the corresponding partition function,
$\overline{Z_{1N}}$, changes only to the extent that $\gamma_a$ is replaced
by a boundary exponent $\gamma_{1a}$, while the exponential growth
has the same $K_a$\cite{vanderzande,note1}. 
Thus, $\overline{Z_N} / \overline{Z_{1N}}$ grows
as $N^{\gamma_a-\gamma_{1a}}$, which is more easy to
extrapolate\cite{seno}.
If one assumes for the quenched free energies $exp(\overline{\log(Z_N)})$
and $exp(\overline{\log(Z_{1N})})$ similar behaviors with exponents
$\gamma$ and $\gamma_1$, respectively,
$\gamma-\gamma_1=\gamma_a-\gamma_{1a}$ should hold  for extrapolated
differences, as long
as annealed and quenched free energies coincide. The temperature 
at which the two 
differences possibly cease to be equal should be identified with $T_d$.
If $T_{a\theta}>T_d$, one must also find $T_{a\theta}=T_{\theta}$ and all
theta exponents in the annealed and quenched problems are the same.
$T_d$ determinations based on checking 
validity of $\gamma-\gamma_1=\gamma_a-\gamma_{1a}$ are much 
more precise than those based on the analysis of exponential growths, 
and reveal of key importance for our investigation.

\begin{figure}[]
\centerline{\hbox{
\epsfysize=17\baselineskip
      \epsffile{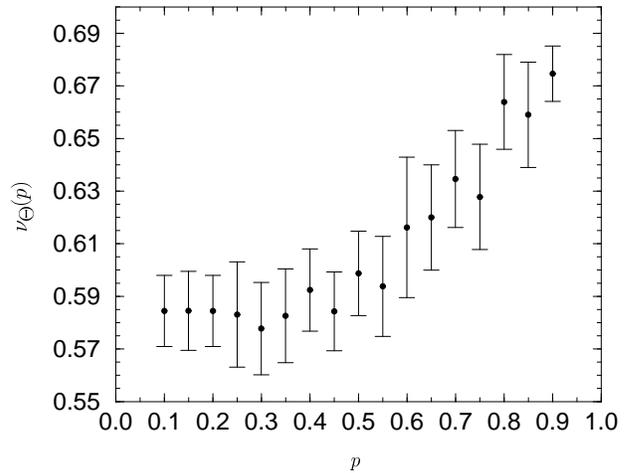}  }}
\caption [Figure]
	 {\protect\footnotesize 
Extrapolated $\nu_{\theta}$ as a function of $p$. For details of these 
and other determinations see ref. \cite{monari}.
	 }  
\label{nu4}
\end{figure}

The basic tool of our approach is the exact determination of all possible
contact maps\cite{lifson} for a polymer of $N$ steps. The contact map in
a given configuration $\omega$ is the set of contacts $\med{ij}\in \omega$.
After all maps have been sorted, since
interactions pertain to contacts, the (entropic) free energy associated
to each map can be evaluated exactly once for all,
and used in order to perform relatively fast averages over the 
disorder affecting the interactions. This latter
averaging could be performed either exactly, or by extensive Monte Carlo (MC)
sampling ( up to $2\times 10^4$ potential configurations)
for the longest chains considered here. Our contact map
algorithm alone can treat chains of length exceeding by at least 4 steps
the maximum length reached in most recent work\cite{vendruscolo}.

\begin{figure}[]
\centerline{\hbox{
\epsfysize=17\baselineskip
      \epsffile{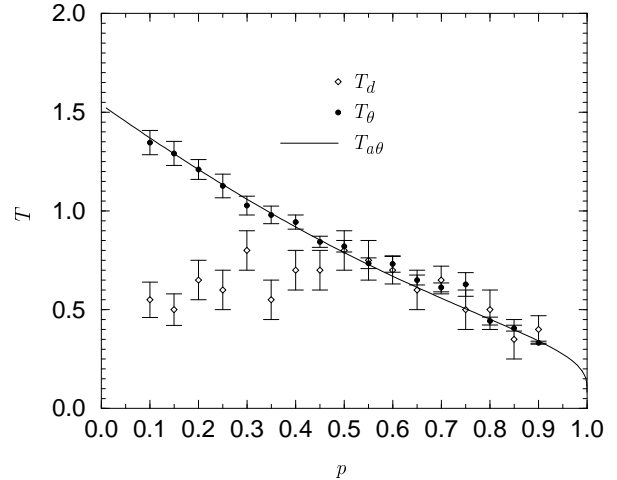}  }}
\caption [Figure]
	 {\protect\footnotesize 
Temperatures playing a role in the problem.
	 }  
\label{temps}
\end{figure}

We determined for $N$ up to $22$ and all $T$ the averages
\begin{equation}
\overline{\med{R^2_N}}=\Sigma_{\{V\}}\Pi_{\med{ij}}P(V_{ij}) \med{R^2_N}\{V\}
\label{2}
\end{equation}
where $\med{R_N^2}\{V\}$ is the thermally averaged 
end-to-end distance for a particular $\{ V \} $.
$\overline{\med{R_N^2}}$ is expected to grow with $N\to \infty$ as a
power law with the three exponent values mentioned above. 
%for $T>T_{\theta}$, $T=T_{\theta}$ and $T<T_{\theta}$.
We expect that effective exponents 
$\nu(M,K,T)=\frac{1}{2}\log  ( 
\overline {\med{R^2_M}}/\overline{\med{R^2_{M-K}}} )
/\log \left ( {M}/({M-K})\right )$, should
interpolate smoothly between values, which, for
increasing $M$, approach the swollen
and the compact $\nu$ exponents, for $T>T_{\theta}$ and $T<T_{\theta}$,
respectively. If the trends of approach are monotonic and opposite in the 
two cases, it is also very likely to find that the various curves
$\nu=\nu(M,K,T)$ bend quite rapidly and 
intersect with each other in a narrow region of the $(T,\nu)$ plane, 
each intersection
representing an approximate determination of the asymptotic $\nu_{\theta}$.
In fact the pattern of $T$-dependence we found for $\nu(M,K,T)$ is of this 
kind for all values of $p$ we tested.
We identified $T_{\theta}$ as the center of the relatively
narrow $T$ range within which the trend of various $\nu(M,K,T)$ changes
from monotonically increasing, to monotonically decreasing with $M$. This
can be done by use of suitable data correlators\cite{monari}. $\nu_{\theta}$
is then determined by extrapolation of 
$\overline{\med{R_N^2}}$ at $T_{\theta}$. 
Figs.\ref{nu4} and \ref{temps} show determinations of $\nu_{\theta}$ 
and $T_{\theta}$, respectively. $T_{a\theta}$ is known with
high precision based on the mapping of the annealed problem onto the
effective homopolymer one ( Fig.\ref{temps}).
$T_{\theta}$ remains remarkably close to $T_{a\theta}$ in
the whole range $0<p<1$. This means that $T_d$ should never overcome 
$T_{\theta}$,
even for very large $p$'s. 
That $T_d>T_{\theta}$ 
should be ruled out is rather plausible, since sequence disorder is very
unlikely to affect the swollen phase.
On the other hand, the
behavior of $\nu_{\theta}$ as a function of $p$ is pretty stable and nicely 
consistent with the homopolymer universality class ($\nu_{\theta}\sim 4/7$)
only for $p\leq0.50 \div 0.55$. For larger $p$'s $\nu_{\theta}$ starts deviating
rather markedly ($\nu_{\theta}\sim 0.64$) from $4/7$ 
and increases
up to $\sim 0.68$ for $p$ close to $1$ . Even if uncertainties do not allow to
identify precisely a different plateau for $p \gsim  0.55$, the
deviation from the homopolymer theta point universality class is very
clear. This evidence is enforced by further results for the
crossover  exponent $\phi_{\theta}$ defined by $\frac{d}{dT}\overline{
\med{R^2_N}}|_{T=T_{\theta}} \sim N^{\phi_{\theta}+2\nu_{\theta}}$
and extrapolated  from our determinations of this derivative
at $T_{\theta}$. 
The pattern is qualitatively similar to that of Fig.\ref{nu4}.
A plateau ($\phi\sim 0.45 $) slightly above the exact homopolymer
$\phi_{\theta}=3/7$ \cite{duplantier} 
is seen also in this case for $p<0.50 \div  0.60$, while
for $p \gsim 0.60$ again clear increasing deviations from this value occur.

\begin{figure}[]
\centerline{\hbox{
\epsfysize=16\baselineskip
      \epsffile{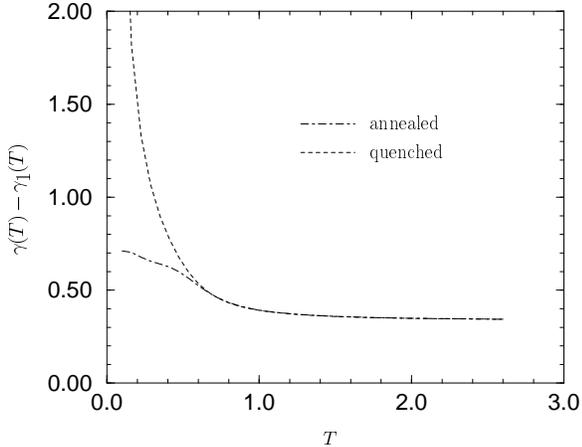}  }}
\caption [Figure]
	 {\protect\footnotesize 
Example of $T_d$ determination for $p=0.6$
	 }  
\label{gamma4}
\end{figure}

The above deviations for $p \gsim  0.50$, combined with $T_{\theta}\sim 
T_{a\theta}$, suggest that, in the upper range of $p$, $T_d$ should be 
very close to,
and possibly coincide with $T_{\theta}$. A strict coincidence is the most 
plausible way in which disorder could affect the universal theta
point properties, while leaving $T_{\theta}=T_{a\theta}$.  
Quite remarkably our determinations of $T_d$ based on extrapolations of
$\gamma-\gamma_1$ and $\gamma_a-\gamma_{a1}$ are pretty consistent with
this conclusion. These exponents are plotted in Fig.\ref{gamma4} 
for $p=0.6$. $T_d$ corresponds to the splitting of the two curves, which
occurs slightly below $T_{\theta}$.
Althought uncertainties are not small, 
$T_d\sim T_{\theta}$ is
clearly suggested by the overall pattern of determinations (Fig.\ref{temps})
for $p \gsim 0.5$, while for $p \lsim 0.5$ one definitely finds
$T_d<T_{\theta}$, consistent with the picture conjectured above.

Frustration is essential
in order to produce the above change of theta universality 
class: distributions $P(V_{ij})$ with support restricted to
attractive potentials did not show such modification. 
Thus, the change from low- to high-$p$ theta regimes could have
analogies with a transition from ferromagnetic to spin glass ordering
\cite{garel}.

Summarizing, we gave strong evidence that a sufficient amount of
frustrated sequence disorder
can be relevant for the heteropolymer behavior in the whole range 
$T \leq T_{\theta}$, determining in particular a universality class
different from that of homopolymers for the theta transition.
This is
strongly suggested by the global consistency of our results
for various exponents and temperatures of both quenched and
annealed problems. The conjecture that
$T_d=T_{\theta}$ in the high frustration regime, is rather natural and 
implies the intriguing possibility that, when frustration is high enough, 
specificity becomes a key factor in the quenched statistics right from
where the heteropolymer starts collapsing. 

The limited accuracy and asymptoticity of our determinations
do not allow a precise conjecture on the nature of the transition
regime in the whole range $0.50 \lsim p < 1$. 
$T_{a\theta}$ approaches $0$ very steeply for $p\to 1^{-}$ 
($T_{a\theta}(p) \sim -1/\log (1-p) $).
Also $T_d$ should approach $0$ for $p\to 1^{-}$ since at all $T$ 
on the $p=1$ line the behavior of the system is that of a SAW, controlled 
by a $T=\infty$ fixed point.
The proximity of this line could be responsible for the increase 
of our $\nu$ estimates ($\nu \sim 0.68$) for $p$ closer to $1$.

\begin{figure}[]
\centerline{\hbox{
\epsfysize=16\baselineskip
      \epsffile{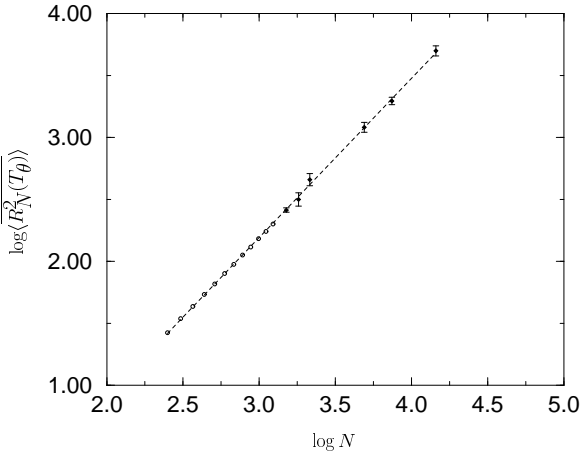}  }}
\caption [Figure]
	 {\protect\footnotesize 
Exact (empty dots) and MC (full) data for $p=0.80$.}  
\label{mc4}
\end{figure}

In order to rule out the possibility that our results could be explained 
just in terms of a very slow crossover to  homopolymer theta behavior, 
due to the presence of the SAW line, 
we spent an exceptional effort in extending to larger $N$ by MC 
methods our determinations of $\overline{\med{R_N^2}}$ for the
particular case $p=0.80$. Experience
\cite{golding,monari} has taught us that, besides the difficulty
of thermal sampling at low $T$, which can, e.g., be solved by application
of multiple Markov chain algorithms
\cite{multiple}, a major limitation of
MC in this field is that quenched averages have to be carried
out over very large ensembles of chain sequences in order to
produce reliable results. This is even more compelling 
when chain specificity 
plays a relevant role. By extensive simulations based on a 
multiple Markov chain method\cite{multiple} (grids of up to $40$ temperatures
and $\sim 600$ different sequences for each $N$) we obtained
extra determinations of $\overline{\med{R_N^2}}$ up to $N=64$.
The log-log plot of Fig.\ref{mc4} confirms very nicely
the trend of the exact enumeration results and clearly excludes a crossover.
On the basis of all data we could estimate $\nu_{\theta}(0.8)=0.64\pm 0.01$
\cite{note3},
which qualifies as our best exponent determination concerning
the expected novel theta universality class at $p \gsim  0.50$.
The value of this exponent is surprisingly very close 
to that appropriate for branched polymers in $2D$ 
\cite{stauffer}.

We expect that the new theta universality class
could be found also in more realistic heteropolymer
models. Polyampholytes with screened interactions 
and nonzero total charge\cite{golding,kardar}
are good candidates. In general the universality 
class at the $\theta$ point should change as soon as frustration 
exceeds a certain threshold.

We thank A. Maritan for stimulating discussions. A.L.S. acknowledges 
partial support from the European Network Contract No. ERBFMRXCT980183.
C.V. thanks the IUAP for support.

\end{multicols}


\begin{references}
\bibitem{note}$\gamma\neq\delta$: the contribution of quadratic terms with
$\gamma=\delta$ in Eq.1 is only responsible for an inessential shift of
$T_{\theta}$.
\bibitem{note1} In the collapsed regime an extra factor $\sim exp(a N^{1/2})$
can enter the partitions. This does not alter our conclusions.
\end{references}
\end{document}